\documentclass[12pt]{article}

\title{Damping of Growth Oscillations in  \\
Molecular Beam Epitaxy:  \\ 
A  Renormalization Group Approach}
\author{Martin Rost\footnote{Email: marost@theo-phys.uni-essen.de.
 Fax: 49-201-183 2120.} and Joachim Krug \\
{\small Fachbereich Physik} \\
{\small Universit\"at GH Essen} \\
{\small D-45117 Essen, Germany}
}

\begin{document}
\maketitle
\begin{abstract} The conserved Sine--Gordon Equation with nonconserved
  shot noise is used to model homoepitaxial crystal growth.
  With increasing coverage the
  renormalized pinning potential changes from strong to weak. This is
  interpreted as a transition from layer--by--layer to rough growth. 
  The associated length and time scales are identified, and found to
  agree with recent scaling arguments. A
  heuristically postulated nonlinear term $\nabla^2 (\nabla h)^2$
  is created under renormalization. 
\end{abstract}

\section{Introduction}
In Molecular Beam Epitaxy (MBE) it is possible to control the amount of
deposited matter through oscillations of the surface roughness, 
which indicate
that the surface grows layer by layer \cite{rheed,cohen,comsa}. 
A simple picture represents
layer-by-layer growth on a high symmetry surface as follows: After
deposition out of the beam atoms diffuse on the surface until they either
meet other diffusing atoms to form a stable island, or get incorporated at
the edge of a previously nucleated island. If most atoms deposited on top of 
an island are assumed to be incorporated into its edge by performing a
downward hop (which implies that the suppression of interlayer
transport by Ehrlich-Schwoebel-barriers \cite{schwoebel} is negligible),
little nucleation occurs 
in the second crystal layer before the first layer is completed. 
Consequently the surface width at layer completion is nearly zero,
after having gone through a maximum
at half filling of the first layer. 
As long as the 
layer-by-layer growth mode persists, the surface morphology exhibits
oscillations with a period given by the monolayer completion time.

In general this scenario is only transient, and the oscillations are damped.
A variety of mechanisms contribute to the damping: The surface may be slightly
miscut \cite{cohen,neave}, or the average beam intensity may be inhomogeneous
\cite{comsa}. However, even in the absence of such (experimentally
unavoidable) imperfections, the stochastic beam fluctuations are
sufficient to destroy the temporal coherence of spatially 
separated regions on the surface. Provided the beam noise is the sole
damping mechanism, it was recently shown that the critical coverage
$\tilde \theta$ at which the oscillations disappear scales with the
ratio of the surface diffusion constant $D_S$ to the deposition flux
$F$ as \cite{wolf,harald}
\begin{equation}
\label{theta}
\tilde \theta \sim (D_S/F)^{\delta}
\end{equation}
with an exponent
\begin{equation}
\label{delta}
\delta = \gamma \frac{4d}{4-d},
\end{equation}
where $d$ denotes the surface dimensionality ($d=2$ for real surfaces) and the exponent $\gamma$
characterizes the dependence of the diffusion length (or typical island size) $\ell_D$ on
$D_S/F$ \cite{submono},
\begin{equation}
\label{gamma}
\ell_D \sim (D_S/F)^\gamma.
\end{equation}
Eq.(\ref{theta}) may therefore be rewritten as
\begin{equation}
\label{theta2}
\tilde \theta \sim \ell_D^{4d/(4-d)} \sim \tilde \ell^d
\end{equation}
where 
\begin{equation}
\label{ltilde}
\tilde \ell \sim \ell_D^{4/(4-d)}
\end{equation}
is the coherence length, 
an estimate of the size of coherently oscillating regions \cite{harald}.

The theory of Ref.\cite{harald} is based on a phenomenological stochastic continuum equation 
for the growing surface. It is not clear {\em a priori} that such a continuum description
would be able to capture the phenomenon of growth oscillations, which is distinctly a
lattice effect. As a first step towards a more complete treatment, in the present work
we therefore {\em perturbatively} include the lattice structure by analyzing the 
driven, conserved sine-Gordon equation  
\begin{equation}
   \label{csg}
   \partial_t h = - K \Delta^2 h - \lambda \Delta (\nabla h)^2 - V \Delta 
   \sin\frac{2 \pi h}{a_\perp}  
    + F + \eta
\end{equation}
for the surface height $h({\bf x},t)$. In this equation 
the constant $F$ denotes the average deposition flux, while the noise $\eta({\bf x},t)$ models
its ``shot noise'' fluctuations. The noise is assumed to be Gaussian with 
mean zero and correlator 
\begin{equation}
\label{noise}
\langle \eta({\bf x},t) \eta({\bf x'},t') \rangle = 2 D \; 
\delta(t-t') \delta({\bf x}-{\bf x'}).
\end{equation}

To motivate the systematic terms on the right hand side of eq.(\ref{csg}), we note that
it can be written in the form of a continuity equation
\begin{equation}
\label{cont}
\partial_t h + \nabla \! \cdot \! {\bf J} = F + \eta 
\end{equation}
reflecting the absence of desorption and defect formation under
typical MBE conditions \cite{villain,lai,adv}. The current is given by
Fick's law, ${\bf J} = - \nabla \rho$, where the quantity
\begin{equation}
\label{rho}
\rho({\bf x},t) = - K \Delta h - \lambda (\nabla h)^2 - V \sin\frac{2
\pi h}{a_\perp}
\end{equation}
can be interpreted as the spatially varying part of a coarse grained
adatom density \cite{villain} or chemical potential \cite{adv}. The
first term incorporates a generalized Gibbs-Thomson effect
\cite{mullins}, according to which the density is enhanced near maxima
and decreased near minima of the surface, while the second term
reflects the dependence of the density on the local vicinality
\cite{harald,adv,korea}. The third term models the lattice potential:
Adatoms are preferably driven to places where they can be incorporated
such that the surface remains at integer multiples of the vertical
lattice constant $a_\perp$. Only the lowest harmonic of the periodic
surface potential is kept, since components of periodicity $a_\perp/2,
a_\perp/3, \dots$ are less relevant (see below and \cite{NG}). It is
necessary to distinguish the vertical ($a_\perp$) and horizontal
($a_\parallel$) lattice constants, since they play very different
roles in the renormalization group calculation. 

The analysis of Ref.\cite{harald} was based on eq.(\ref{csg}) with
$V=0$. Here we show that, by explicitly including the lattice
potential, the characteristic time and length scales
(\ref{theta},\ref{ltilde}) emerge naturally in the renormalization
group (RG) flow equation of the potential strength $V$. Moreover the
$(\nabla h)^2$ nonlinearity in (\ref{rho}) is seen to be generated
under renormalization, thus relieving us from the task of postulating
its microscopic origin; we may set $\lambda = 0$ microscopically. 
A similar scenario is valid for the nonconserved Sine--Gordon model
\cite{Spohn}. These results are obtained by applying the
Nozi\`eres-Gallet RG scheme \cite{NG} to eq.(\ref{csg}). An RG
analysis of (\ref{csg}) was previously presented by Tang and
Nattermann \cite{tn}, however these authors considered separately the
cases $V = 0$, $\lambda \neq 0$ and $V \neq 0$, $\lambda = 0$ and thus
were not able to address the generation of $\lambda$ from the lattice
potential; in addition, our analysis includes explicitly the flat
initial condition of the surface, which is essential for describing
transient behavior.

Since the interpretation of the RG results depends crucially on 
relating the ``mesoscopic'' coefficient $K$ to the microscopic length
scale $\ell_D$, the next section will address this issue within the
framework of the linear equation ($V = \lambda = 0$). It turns out
that the mere existence of a vertical lattice constant $a_\perp$ is
sufficient for the nontrivial time and length scales
(\ref{theta},\ref{ltilde}) to emerge from the continuum theory, even
if this scale has no dynamical effect (i.e., $V = 0$)
\cite{iff97}. The full problem with $V \neq 0$ is treated in the
following sections. After briefly recalling the RG scheme, the
renormalization of the conserved Sine--Gordon Equation in $2 \! + \!
1$ dimensions is carried out in Section 3. The RG--flow equations for
the parameters in Eq.\ (\ref{csg}) are interpreted in Section 4, and
the extension to general dimensionalities and general relaxation
mechanisms is briefly addressed. Some conclusions are given in Section
5.

\section{Length scales in the linear theory} 
\label{Linear}

On the most basic level, the growing surface morphology evolves in
response to the competition between disordering beam fluctuations and
smoothening surface diffusion. The simplest continuum theory that
incorporates both effects is the linearization of (\ref{csg}),
\begin{equation}
\label{mullins}
\partial_t h = - K \Delta^2 h + F + \eta. 
\end{equation}
As was mentioned in Section 1, the coefficient $K$ arises from an
expansion of the local adatom density in the surface curvature $\Delta
h$. Under near-equilibrium conditions, it would therefore be expected
to be given by the product of the surface stiffness and the adatom
mobility \cite{mullins,jkwagner}. However, far from equilibrium other
processes may contribute to, and in fact dominate $K$
\cite{villain,politi1}. In particular, it has been suggested
\cite{politi2} that random island nucleation produces a contribution
\begin{equation}
\label{K}
K \sim F \ell_D^4,
\end{equation}
but the underlying microscopic mechanism is not known. In the
following we show how this relation follows from a simple
reinterpretation of (\ref{mullins}) in the presence of a finite
vertical lattice constant $a_\perp$.

The straightforward solution of (\ref{mullins}) \cite{adv} shows that,
starting from a flat substrate at time $t=0$, after time $t$ surface
correlations have developed up to a scale
\begin{equation}
\label{xi}
\xi(t) \approx (K t)^{1/4}
\end{equation}
and the surface width grows as 
\begin{equation}
\label{W}
W(t) \approx (D/K)^{1/2} \xi(t)^\zeta
\end{equation}
in dimensionalities $d < 4$, where
\begin{equation}
\label{zeta}
\zeta = \frac{4-d}{2}
\end{equation}
is the roughness exponent of the linear equation \cite{adv}. 

Together with the average growth rate $F$ the presence of the vertical
lattice constant induces a fundamental time scale, the monolayer
completion time
\begin{equation}
\label{tauML}
\tau_{\rm ML} = a_\perp/F.
\end{equation}
Setting $t = \tau_{\rm ML}$ in (\ref{xi}) we obtain a corresponding
lateral length scale $\xi(\tau_{\rm ML})$, the scale on which lateral
structure has developed after deposition of one monolayer. Clearly it
is very natural to identify this scale with the diffusion length
$\ell_D$, and hence the coefficient $K$ in (\ref{xi}) can be
identified as 
\begin{equation}
\label{K2}
K \approx \ell_D^4/\tau_{\rm ML} = a_\perp^{-1} F \ell_D^4
\end{equation}
in accordance with (\ref{K}). The correlation length (\ref{xi}) can
then be expressed in terms of the coverage $\theta = F t / a_\perp$ as
\begin{equation}
\label{xild}
\xi(t) \approx \ell_D \theta^{1/4}.
\end{equation}

To see how the scaling laws (\ref{theta}) and (\ref{ltilde}) can be
obtained along similar lines, note first that for shot noise the noise
strength $D$ is proportional to the beam intensity $F$, 
\begin{equation}
D \approx a_\perp a_\parallel^d F.
\end{equation}
Thus (\ref{W}) takes the form
\begin{equation}
\label{W2}
W \approx a_\perp (a_\parallel/\ell_D)^{d/2} \theta^{(4-d)/8}.
\end{equation}
If one now postulates, plausibly \cite{harald}, that the lattice
effects disappear when the roughness due to long wavelength
fluctuations becomes comparable to $a_\perp$, the characteristic
coverage $\tilde \theta$ can be defined through $W(\tilde \theta)
\approx a_\perp$ and is given precisely by eqs. (\ref{theta}) and
(\ref{delta}).

These considerations may be viewed as a zeroth order assessment of
lattice effects, to be justified by the systematic calculation
provided in the remainder of the paper. They easily extended to
general linear equations with a dynamical exponent $z$ \cite{adv}, in
which case one finds \cite{harald}
\begin{equation}
\label{delta2}
\delta = \gamma \frac{zd}{z-d}
\end{equation}
for $z > d$.

\section{Renormalization Group Analysis}

\subsection{The Renormalization Scheme of Nozi\`eres and Gallet}

Because of the structure of Equation (\ref{csg}) we use an approach
which is suitable for general forms of the nonlinearity. It was
introduced by Nozi\`eres and Gallet for the dynamical renormalization
of the Sine-Gordon equation to obtain the roughening transition
\cite{NG}. A detailed presentation can be found in their work, which
we recall briefly.

Consider a Langevin equation 
\begin{equation}
  \label{ngeq}
  \partial_t h = {\cal L} h + {\cal N}(h) + \eta
\end{equation}
with a linear part ${\cal L} h$, a nonlinear term ${\cal N}(h)$ and
Gaussian noise $\eta$ with mean zero and correlator $\langle \eta({\bf
  k},t) \eta({\bf k'},t') \rangle = 2 |{\bf k}|^{2\mu} D \delta(t-t') 
\delta({\bf k}+{\bf k'}) \theta(|{\bf k}|-\Lambda)$. The cutoff
$\Lambda \! \equiv \! 1/a_\parallel$ is introduced to model the
lateral lattice structure which does not allow for fluctuations on
scales smaller than the horizontal lattice constant $a_\parallel$.
In the following $a_\parallel$ will only appear in the cutoff
$\Lambda$, hence we can disregard the distinction between
$a_\parallel$ and $a_\perp$ and set $a_\perp = a$.

Two types of noise can be considered: Either volume conserving noise,
which corresponds to the case $\mu \! = \! 1$ (``diffusion noise''
\cite{wolf}), or nonconserving ``shot'' noise $\mu \! = \! 0$.
Here we focus on the nonconserved contribution, $\mu = 0$, which
always dominates on scales larger than the diffusion length $\ell_D$
\cite{harald,wolf}. A study of the conserved case has been presented
in Ref.\cite{sun}.

Renormalization of Equation (\ref{ngeq}) is performed in the following
way:
\begin{itemize}
\item We average over the short wave components $\delta \eta$ of the
  noise. In {\bf k}--space $\delta \eta$ is nonzero only for modes
  {\bf k} with $(1 - dl)\Lambda < |{\bf k}| \leq \Lambda$. Define the
  averaged or coarse grained field $\bar h \equiv \langle h
  \rangle_{\delta \eta}$.
\item Equation (\ref{ngeq}) is split in two; one for the coarse grained
  field
\begin{displaymath}
  \partial_t \bar h = {\cal L} \bar h + \langle {\cal N}(\bar h +
  \delta h) \rangle_{\delta \eta}+ \bar \eta 
\end{displaymath}
  and a second one for the difference $\delta h \equiv h - \bar h$. One now
  takes an approximation of $\langle {\cal N}(\bar h + \delta
  h)\rangle_{\delta \eta}$ 
  in terms of the (hopefully all) relevant operators appearing in
  Equation (\ref{ngeq}). For this one calculates $\delta h$
  (respectively its correlation functions) perturbatively in the
  nonlinearity ${\cal N}$. Since ${\cal N}$ is not of a simple polynomial
  form, a Rayleigh--Schr\"odinger expansion is used -- the only one
  feasible one, albeit poorly controlled.
\item Time, lateral and vertical length are rescaled with different
  exponents: ${\bf x} \to (1 - dl) {\bf x}$, $h \to (1 - \zeta \; dl)
  h$ and $t \to (1 - z \; dl)t$. This causes a corresponding rescaling
  of the terms in (\ref{ngeq}) yielding the RG flow equations for
  ${\cal L}$ and ${\cal N}$.
\end{itemize}
The detailed application of these steps to Equation (\ref{csg}) is
the subject of the next section.

\subsection{Application to the Conserved Sine--Gordon Equation}
Consider Equation (\ref{csg}) in a frame moving with the average
growth speed $F$
\begin{equation}
  \label{csg2}
  \partial_t h = - K \Delta^2 h - V \; \Delta \sin \frac{2 \pi}{a}
  \biggl( h - Ft \biggr) + \eta.
\end{equation}
Epitaxial growth starts on an atomically flat surface, so at time $t
\! = \! 0$ the initial configuration is $h \! \equiv \!
0$. Fluctuations are caused by the noise at {\em later} times. This
will play an important role in the interpretation of our results.

We neglect  the perturbative contribution of the $- \lambda \Delta
(\nabla h)^2$--term. As argued in
the Introduction on a microsopic level it is absent.
It is generated by the lattice potential and the driving force $F$ 
to order $V^2$. Although it is a relevant operator at the linear fixed point,
we expect it to be negligible on small and intermediate scales, as
long as the lattice potential contributes to renormalization of $K$
and $\lambda$. The same is observed for the $(\nabla
h)^2$--nonlinearity in the nonconserved Sine--Gordon Equation
\cite{Spohn}. 

Averaging over $\delta \eta$ leads to two coupled equations, where we
expand to lowest order in the infinitesimal quantity $\delta h$
\begin{eqnarray}
  \label{coupled}
  \partial_t \bar h & = & - K \; \Delta^2 \bar h - V \Delta \left[ \sin
  \frac{2 \pi}{a} \biggl( \bar h - Ft \biggr) \left( 1 - \frac{2
  \pi^2}{a^2} \langle 
  \delta h^2 \rangle \right) \right] + \bar \eta \nonumber \\
  \partial_t \delta h & = & - K \; \Delta^2 \delta h - V \Delta \left[ \cos
  \frac{2 \pi}{a} \biggl( \bar h - Ft \biggr) \; \; \frac{2 \pi}{a} \;
  \; \delta h \right] + \delta \eta .
\end{eqnarray}
The second equation of (\ref{coupled}) is solved by a
Rayleigh--Schr\"odinger perturbation ansatz
\begin{eqnarray}
  \label{perturb}
  \delta h^{(0)}({\bf x},t) & = & \int d^d x' \int_0^t dt' \; G({\bf x -
  x'},t-t') \; \delta \eta({\bf x'},t') \\
  \delta h^{(1)}({\bf x},t) & = & - \frac{2 \pi V}{a} \int d^d x'
  \int_0^t dt' \; G({\bf x - x'},t-t') \; \times \nonumber \\
  & & \; \; \; \Delta' \left[ \cos \frac{2 \pi}{a} \biggl(\bar h({\bf
  x'},t') - Ft' \biggr)  \; \delta
  h^{(0)}({\bf x'},t')  \right]. \nonumber
\end{eqnarray}
In ${\bf k}$--space the linear propagator $G$ is given by $\exp -K
(t-t') |{\bf k}|^4$. 
In the sequel we will use the notation $\bar h', \eta', \dots$ when
the argument is the integration variable $({\bf x'},t')$ and
unprimed symbols for quantities at $({\bf x},t)$. The corresponding
derivatives (Laplace operators) are denoted $\Delta$ and $\Delta'$ to
mark this difference. For the convolution integral $\int d^d x'
\int_0^t dt'$ the shorthand $\int$ is used. 

Insertion of $\langle \delta h^2 \rangle \! = \!  \langle {\delta
h^{(0)}}^2 \rangle + 2 \langle \delta h^{(0)} \delta h^{(1)} \rangle
+ O(V^2)$ in the first equation of (\ref{coupled}) generates the
corrective terms for $\bar h$ to order $V^2$.
\begin{equation}
  \langle \delta h^2 \rangle = \langle {\delta h^{(0)}}^2 \rangle -
  \frac{4 \pi V}{a} \int (\Delta' G) \; \cos \frac{2 \pi}{a} \biggl(
  \bar h' - F t' \biggr) \; C,
\end{equation}
where $C \equiv \langle \delta h^{(0)} \delta {h^{(0)}}' \rangle $ is
the unperturbed correlator. Its Fourier transform is given by $e^{-K
(t-t') |{\bf k}|^4} D/(K|{\bf k}|^4) \left[ 1 - e^{-2Kt|{\bf
k}|^4} \right]$. 

The lowest order term $\langle {\delta h^{(0)}}^2 \rangle$ is a
constant which multiplies the sine potential, yielding the correction
\begin{equation}
  \label{dV}
  \delta V = - \frac{4 \pi^3 D}{K a^2 \Lambda^2} \left( 1 - e^{-2 K
  \Lambda^4 t} \right) \; dl \; V.
\end{equation}
In general dimension $2 \pi/\Lambda^2$ is replaced by 
$S_d \Lambda^{d-4}$, where $S_d$ is the surface area of the $d$-dimensional
unit sphere.

To handle the next order contribution due to $\langle \delta h^{(0)}
\delta h^{(1)} \rangle$,
\begin{displaymath}
  - \frac{8 \pi^3 V^2}{a^3} \; \; \Delta \int \sin \frac{2 \pi}{a}
  \biggl( \bar h \! - \! Ft \biggr) \; \cos \frac{2 \pi}{a} \biggl(
  \bar h' \! - \! Ft' \biggr) \; ( \Delta' G) \; C,
\end{displaymath}
we split the product $\sin \alpha \cos \beta = (\sin(\alpha\!+\!\beta) + 
\sin(\alpha\!-\!\beta))/2$ and discard the term with $\sin(\alpha \! +
\! \beta)$. It would create higher harmonics of the lattice potential,
which are less relevant (see Eq.\ (\ref{dV})) than the fundamental. In
the remaining expression we split into terms with $\bar h$ and $Ft$
\begin{eqnarray}
  - \frac{4 \pi^3 V^2}{a^3} & \Delta & \int \Biggl[ \sin \frac{2
  \pi}{a} F(t'-t) \; \cos \frac{2 \pi}{a} \biggl( \bar h - \bar h'
  \biggr) \; (\Delta' G) \; C \nonumber \\
 & + & \cos \frac{2 \pi}{a} F(t'-t) \; \sin \frac{2 \pi}{a} \biggl(
  \bar h - \bar h' \biggr) \; (\Delta' G) \; C \Biggr]. \nonumber
\end{eqnarray}
Expanding terms with $\bar h \! - \! \bar h'$ in powers of $\xi \!
  \equiv \! {\bf x' \! - \! x}$ and taking care of the symmetries when 
integrating over $d^2\xi$ we get
\begin{eqnarray}
\label{corrections}
= &  &  \frac{4 \pi^5 V^2}{a^5} \; \biggl[ \Delta (\nabla \bar h)^2
\biggr] \; \int \! \sin \left[ \frac{2 \pi}{a} F(t'-t) \right]
(\Delta' G) \; C \; \xi^2 \nonumber \\ 
 & - & \frac{2 \pi^4 V^2}{a^4} \; \biggl[ \Delta^2 \bar h \biggr] \; \int
 \! \cos \left[ \frac{2 \pi}{a} F(t'-t) \right] (\Delta' G) \; C \;
 \xi^2. 
\end{eqnarray}
Up to now we have expanded the averaged nonlinearity $\langle
{\cal N}(h) \rangle_{\delta \eta}$ to order $V^2$ in projections onto
the relevant operators.

The last step consists in rescaling space and time. To examine the
behavior close to the linear fixed point $V \! = \! 0, \lambda \!
= \! 0$ we choose its scaling exponents, $z \! = \! 4$ and $\zeta
\! = \! 2 \! - \! d/2 \! = \! 1$ in two dimensions (see (\ref{zeta})). 
Accordingly the growth rate $F$ and the lattice constants $a_\perp$ 
and $a_\parallel$ are rescaled as
$dF/dl \! = \! (z \! - \! \zeta)F$, $da_\perp/dl \! = \!
- \zeta a_\perp$ and $da_\parallel/dl \! = \!
- a_\parallel$.

\subsection{Flow equations}

The RG flow will be examined in terms of dimensionless quantities,
in which the effect of the trivial rescaling has been eliminated.
As elementary length scales we use
the vertical and horizontal lattice constants 
$a_\perp$ and $a_\parallel$, and the basic time scale is given by the
monolayer completion time $\tau_{\rm ML} = a_\perp/F$. Thus time
is measured through the coverage $\theta = t/\tau_{\rm ML}$, and the
coefficients $K$, $V$ and $\lambda$ appearing in (\ref{csg}) are
replaced by the expressions $${\cal K} \! \equiv \! \frac{K a_\perp}{F
a_\parallel^4} = (\ell_D/a_\parallel)^4 $$ $$ U = \frac{V}{F
a_\parallel^2} $$ $$ L = \frac{\lambda a_\perp^2}{F a_\parallel^4}. $$
Due to the conserved form of the equation of motion (\ref{csg}), the
noise strength $D$ is not renormalized \cite{lai,tn}, and need not be
considered further.

Using the scaling parameter $\kappa \! \equiv \! \exp 4l$, the flow
equations then take the form
\begin{eqnarray}
  \label{flow}
 ({\rm i}) \; \; \; \; \; \;   4 \kappa \; \frac{dU}{d\kappa} & = & 
  - 4 \pi^3 \; \frac{\sqrt{\kappa}}{\cal K} \left( 1 - e^{-2 {\cal K}
  \theta/\kappa} \right)  U, \nonumber \\
 ({\rm ii}) \; \; \; \; \; \; 4 \kappa \; \frac{d{\cal K}}{d\kappa} & = &
  \frac{\sqrt{\kappa} \; U^2}{\cal K} \; \; g({\cal K}/\kappa,\theta), \\
 ({\rm iii}) \; \; \; \; \; \;4 \kappa \;  \frac{dL}{d\kappa} & = &
  \frac{\sqrt{\kappa} \; U^2}{{\cal K}} \; \; f({\cal K}/\kappa,\theta). \nonumber
\end{eqnarray}
The functions $f$ and $g$ depend on the integrals in Eq.\
(\ref{corrections}) and are given in the appendix. 
Equations (\ref{flow}) are the central result of this paper. The following sections
are devoted to the discussion of their physical content. 

\section{Interpretation of the RG Results}

As in the case of the roughening transition \cite{NG}, the
renormalization of $U$ determines the relevance of the lattice on
large scales. In the present situation the lattice potential always
decreases under renormalization (eq.(\ref{flow}(i))). Indeed, since in
the absence of the lattice potential the roughness exponent
(\ref{zeta}) $\zeta > 0$ in two dimensions, the lattice becomes
irrelevant asymptotically; a roughening transition is possible only if
$\zeta = 0$, such as for eq.(\ref{csg}) subject to conserved noise
\cite{sun}. Nevertheless on finite length or time scales $U$ remains
finite, and its dependence on $\theta$ and $\kappa$ may be used to
describe the transition from (lattice-dominated) layer-by-layer growth
to rough, continuous growth. It can be shown that for small $U$ the
amplitude of the surface width oscillations is proportional to $U$;
thus the value of $U$ on a given time or length scale is a direct
measure of the observable signatures of layer-by-layer growth.  

In the following we assume that the rate of particle deposition is
small compared to the diffusion rate, which is true for typical MBE
conditions and implies that the dimensionless stiffness parameter
${\cal K} \! \gg \! 1$. The renormalization of ${\cal K}$ due to the
lattice potential, as expressed by eq.(\ref{flow}(ii)), can then be
disregarded, and ${\cal K}$ becomes a constant. This decouples the
flow equation for the lattice potential $U$ and allows for a
straightforward solution, which can be used to extract the damping
time $\tilde \theta$ and the coherence length $\tilde \ell$. In
Section \ref{Lambda} the generation of the $\Delta (\nabla h)^2$
nonlinearity, as described by eq.(\ref{flow}(iii)), will be discussed.

\subsection{Damping time}

The asymptotic ($\kappa \to \infty$) value of the lattice potential
$U$ is obtained from eq.(\ref{flow}(i)) as
\begin{equation}
\label{dampU}
U_\infty  = e^{-I(\theta)} U_0
\end{equation}
with 
\begin{equation}
\label{Itheta}
I(\theta) = \frac{\pi^3}{\cal K} \int_1^\infty \frac{d\kappa}{\sqrt{\kappa}} 
(1 - e^{-2 {\cal K} \theta/\kappa}) \approx \sqrt{8 \pi^7 \theta/{\cal K}}
\end{equation}
in the relevant regime ${\cal K} \theta \gg 1$. Thus writing $I(\theta) = \sqrt{\theta/\tilde 
\theta}$ the characteristic coverage is obtained as
\begin{equation}
\label{thetaRG}
\tilde \theta = \frac{\cal K}{8 \pi^7} \sim (\ell_D/a_\parallel)^4,
\end{equation}
in agreement with (\ref{theta2}) for $d=2$. Moreover, since in this
case the surface width of the linear theory is proportional to
$(\theta/\tilde \theta)^{1/4}$ (see Section \ref{Linear}), we see that
the decay of the lattice potential is of the form 
\begin{equation}
\label{U2}
U_\infty/U_0 \approx \exp[-C W^2(\theta)]
\end{equation}
where $C > 0$ is a constant. This behavior has been found to describe
the decay of oscillation amplitudes in layer-by-layer growth in
numerical simulations, and can be derived assuming a discrete
probability distribution of the heights taking at each possible height
the value of the corresponding continuous Gaussian distribution
\cite{evans}.

It is instructive to extend these results to other dimensionalities
and linear relaxation mechanisms with a general dynamic exponent
$z$. For general $z$, the natural scaling variable is $\kappa = \exp z
l$. Then the only qualitative change of the expressions discussed
above is that the algebraic part of the integrand in (\ref{Itheta})
becomes $\kappa^{2 \zeta/z - 1}$ instead of $\kappa^{-1/2}$. To see
this, note that the only quantity changing under rescaling in the
time-independent part of the correction $\delta V/V$ in eq.(\ref{dV})
is the vertical lattice spacing $a_\perp \sim
\kappa^{-\zeta/z}$. Thus, for $\zeta < 0$ the integral (\ref{Itheta})
converges even when $\theta \to \infty$, implying that the surface
remains smooth. For $\zeta > 0$ the integral becomes 
\begin{equation}
I(\theta) \sim {\cal K}^{-1} ({\cal K} \theta)^{2 \zeta /z} \sim W^2,
\end{equation}
showing that (\ref{U2}) remains valid in the general case. Writing
$I(\theta) = (\theta/\tilde \theta)^{2 \zeta /z}$, the characteristic
coverage is of the order 
\begin{equation}
\tilde \theta \sim {\cal K}^{z/2 \zeta - 1} = {\cal K}^{d/(z-d)},
\end{equation}
where in the last step the scaling relation $z = d + 2 \zeta$ for
linear growth equations \cite{adv} has been used. This becomes
equivalent to (\ref{delta2}) by noting that (\ref{K2}) is replaced by
$K \approx \ell_D^z/\tau_{\rm ML}$ for general $z$.

\subsection{Coherence length}

For $\theta \to \infty$ the solution of the flow equation
(\ref{flow}(i)) at finite $\kappa$ is 
\begin{equation}
\label{Ukappa}
U(\kappa) = \exp[- (2 \pi^3/{\cal K}) (\sqrt{\kappa} - 1)] U_0.
\end{equation}
If the renormalization is stopped at a lateral scale $L_\parallel =
\kappa^{1/4} a_\parallel$, the remaining value of the lattice
potential may therefore be written, for $L_\parallel/a_\parallel \gg
1$, as 
\begin{equation}
\label{UL}
U(L_\parallel)/U_0 = e^{-(L_\parallel/\tilde \ell)^2}
\end{equation}
with 
\begin{equation}
\tilde \ell = a_\parallel \sqrt{\frac{\cal K}{2 \pi^3}} \sim
\frac{\ell_D^2}{a_\parallel}, 
\end{equation}
in agreement with the expression (\ref{ltilde}) for the coherence length. 
If the system is smaller than $\tilde \ell$
the lattice potential remains relevant, the surface remains smooth and
growth oscillations will be present for
all times \cite{harald}.

As in the previous section, these considerations can be generalized to
arbitrary $z$ and $d$. Then (\ref{UL}) becomes $U(L_\parallel)/U_0 =
\exp[-(L_\parallel/\tilde \ell)^{2 \zeta}]$ with 
\begin{equation}
\tilde \ell \sim {\cal K}^{1/2 \zeta} = {\cal K}^{1/(z-d)} \sim
(\ell_D/a_\parallel)^{z/(z-d)}
\end{equation}
in accordance with (\ref{delta2}) and Ref.\cite{harald}.

\subsection{Generation of the conserved KPZ nonlinearity}
\label{Lambda}

We now focus on the flow equation (\ref{flow}(iii)), which describes
the generation of the nonlinear term of the conserved
Kardar-Parisi-Zhang \cite{kpz} (CKPZ) equation, $\Delta (\nabla h)^2$,
through the interplay of the lattice potential $V$, the growth rate
$F$ and the effective stiffness $K$. We consider the stationary regime
$\theta \to \infty$, and again assume that ${\cal K}$ is large, so
that its renormalization can be neglected. Setting $L = \lambda = 0$
at the microscopic scale, the solution of the flow equation
(\ref{flow}(iii)) then reads
\begin{equation}
\label{Lkappa}
L(\kappa) = \frac{U_0^2}{4 {\cal K}} \int_1^\kappa
dx \; x^{-1/2} \; \exp[-(4 \pi^3/{\cal K})(\sqrt{x}-1)] f({\cal K}/x),
\end{equation}
where the solution (\ref{Ukappa}) for the flow of $U$ has been used.
For $\kappa \to \infty$, eq.(\ref{Lkappa}) tends to a finite limit,
which for ${\cal K} \to \infty$ has the simple form
\begin{equation}
\label{Linfty}
L_\infty = 2 \pi^2 U_0^2. 
\end{equation}
The numerical solution of the full set (\ref{flow}) of coupled flow
equations shows that the limiting value (\ref{Linfty}) is attained for
${\cal K} \geq 10^8$, corresponding to a diffusion length $\ell_D =
100 a_\parallel$. In the range $10^4 \leq {\cal K} \leq 10^8$ the
nonlinear coupling $L\infty$ is positive and increases with increasing
${\cal K}$, while for smaller values of ${\cal K}$ it is negative.

In terms of the bare parameters of the original equation, the relation
(\ref{Linfty}) implies that, on large scales and for large diffusion
lengths, the CKPZ coefficient $\lambda$ is positive and of the form
\begin{equation}
\label{lambda}
\lambda \approx \frac{V_0^2}{F a_\perp^2}.
\end{equation}
It is instructive to compare this to a heuristic estimate of $\lambda$,
based on Burton-Cabrera-Frank (BCF) theory \cite{harald,adv}. 
According to BCF \cite{bcf}, the adatom density on a terrace can be
computed by solving a steady state diffusion equation with sinks at
the surface steps. The density decreases with decreasing step distance
or increasing tilt, and becomes independent of the tilt when the step
distance is of the order of the diffusion length $\ell_D$. The
coefficient $\lambda$ of the leading order expansion (\ref{rho})
around a singular surface is then positive\footnote{For a vicinal
surface growing in the step flow mode one can show that $\lambda < 0$,
see \cite{korea}.} and given by
\begin{equation}
\label{lambdabcf}
\lambda_{\rm BCF} \approx \frac{F \ell_D^4}{a_\perp^2}.
\end{equation}

To identify eqs.(\ref{lambda}) and (\ref{lambdabcf}) we would need to
require that the bare pinning potential $V_0$ depends on the diffusion
length as $V_0 \approx F \ell_D^2$, and hence the dimensionless
potential strength is
\begin{equation}
\label{U0bcf}
U_0 \approx (\ell_D/a_\parallel)^2 = \sqrt{\cal K} \gg 1,
\end{equation}
which is clearly inconsistent with our perturbative treatment of the 
potential. Thus, the expressions (\ref{lambda}) and (\ref{lambdabcf})
are {\em not} equivalent, but rather correspond to different limiting
situations: Our calculation is an expansion for small $U_0$ and fixed
(large) ${\cal K}$, while the BCF picture assumes perfect crystal
planes, which would be formally represented by taking $U_0 \to \infty$
at fixed ${\cal K}$. Writing $L_\infty = L_\infty(U_0, {\cal K})$ we
have shown that $L_\infty = U_0^2 \; \phi({\cal K})$ for $U_0 \to 0$,
where $\phi$ is an increasing function of ${\cal K}$, and the BCF
argument indicates that $L_\infty \sim {\cal K}$ for $U_0 \to
\infty$. Clearly the latter regime is not accessible by our
method. Nevertheless it is remarkable that the CKPZ coefficient
emerges from the RG calculation with the correct sign and the correct
qualitative dependence on the diffusion length.

\section{Conclusions}
In this work we have derived renormalization group flow equations for the
conserved Sine--Gordon Equation with nonconserved shot noise. They were
used to model the crossover in homoepitaxy from layer--by--layer
growth to rough growth. The crossover can be quantitatively
characterized by a characteristic {\em layer coherence length} $\tilde
\ell$ and an associated coverage $\tilde \theta$, which is a measure 
of the number of growth oscillations that can be observed under
optimal growth conditions (that is, in the absence of miscut or beam
inhomogeneity). The dependence of these length and time scales on
growth parameters is in agreement with dimensional analysis and
numerical simulations \cite{harald}.

Our approach also predicts the presence of a nonlinear term of the
form $\Delta (\nabla h)^2$ on large scales, which was suggested
previously on heuristic grounds \cite{villain,lai}. By power counting
it is seen to be relevant in the long time limit and it has nontrivial
effects on the scaling behavior \cite{Janssen}.

Two extensions of this work seem to be possible within the conserved
Sine--Gordon ansatz: First, renormalization of a tilted surface (as
performed for the original Sine--Gordon model in Ref.\cite{NG}) should
clarify the influence of a small miscut on the damping of growth
oscillations \cite{cohen}. Second, and more ambitiously, the
implementation of an Ehrlich--Schwoebel--effect \cite{schwoebel} may
provide a systematic approach to computing the surface current induced
by step edge barriers \cite{villain,adv,kps}, and thus contribute to
understanding the transition from layer-by-layer growth to a
coarsening mound morphology \cite{adv,rost97}.

{\bf Acknowledgements:} We thank H.\ Kallabis for helpful
discussions. This work was supported by Deutsche
Forschungsgemeinschaft within SFB 237 {\em Unordnung und grosse
Fluktuationen}.

\appendix
\section{Appendix: Scaling Functions}
The functions used in the flow equations (\ref{flow},i) and
(\ref{flow},ii) are
\begin{eqnarray}
g({\cal K}/\kappa,\theta) & = &  8 \pi^5 \; \; \frac{ 1 - e^{-2{\cal K}
  \theta/\kappa}}{[ ({\cal K}/\kappa)^2 + \pi^2]^3} \nonumber \\
& &  \Biggl[ {\cal
  K}/\kappa \biggl( -({\cal 
  K}/\kappa)^4 - 4 \pi^2 ({\cal K}/\kappa)^2 + 5 \pi^4 \biggr)
  \nonumber  \\ 
 & & + \Biggl( B \; \cos(2\pi\theta) + A \; \sin(2\pi\theta)  \Biggr) \; \; \; e^{-2{\cal K}
  \theta/\kappa}  
 \Biggr] \nonumber \\
f({\cal K}/\kappa,\theta) & = & 16 \pi^6 \; \; \frac{ 1 - e^{-2{\cal K}
  \theta/\kappa}}{[ ({\cal K}/\kappa)^2 + \pi^2]^3} \nonumber \\
& & 
  \Biggl[ \pi \biggl( - ({\cal K}/\kappa)^4 
- 8 \pi^2 ({\cal K}/\kappa)^2 +  \pi^4 \biggr) \nonumber \\
 & & + \Biggl( - A \; \cos(2\pi\theta) \; \; + B \; \sin(2\pi\theta)  \Biggr) \; \;
  \;  e^{-2{\cal K} \theta/\kappa}
 \Biggr] \nonumber 
\end{eqnarray}
with the polynomials
\begin{eqnarray}
A & = & 4 \pi \theta^2 ({\cal K}/\kappa)^6  + (\pi - 8 \pi^3 \theta^2) ({\cal K}/\kappa)^4
  \nonumber \\
& & + 8 \pi^3 \theta ({\cal K}/\kappa)^3  + (8 \pi^3 - 4 \pi^5 \theta^2) ({\cal K}/\kappa)^2 \nonumber \\
& & +  8 \pi^5
\theta {\cal K}/\kappa - \pi^5  \nonumber \\
B & = & -4 \theta^2 ({\cal K}/\kappa)^7 + 4 
 \theta ({\cal K}/\kappa)^6 + (1 - 8 \pi^2 \theta^2) ({\cal
   K}/\kappa)^5 \nonumber \\
& & + 16 \pi^2 \theta ({\cal K}/\kappa)^4  + 4 (\pi^2 - \pi^4 \theta^2)
({\cal K}/\kappa)^3 + 12
\pi^4 \theta  ({\cal K}/\kappa)^2 \nonumber \\
& & - 5 \pi^4 {\cal K}/\kappa. \nonumber
\end{eqnarray}

\setlength{\unitlength}{0.15bp}
\begin{picture}(2520,2160)(0,0)
\Large
\put(2460,180){\makebox(0,0){$10^{10}$}}
\put(2140,180){\makebox(0,0){$10^9$}}
\put(1820,180){\makebox(0,0){$10^8$}}
\put(1500,180){\makebox(0,0){$10^7$}}
\put(1180,180){\makebox(0,0){$10^6$}}
\put(860,180){\makebox(0,0){$10^5$}}
\put(540,180){\makebox(0,0){$10^4$}}
\put(480,2040){\makebox(0,0)[r]{0.2}}
\put(480,1692){\makebox(0,0)[r]{0.15}}
\put(480,1344){\makebox(0,0)[r]{0.1}}
\put(480,996){\makebox(0,0)[r]{0.05}}
\put(480,648){\makebox(0,0)[r]{0}}
\put(480,300){\makebox(0,0)[r]{-0.05}}
\put(1660,90){\makebox(0,0){$\cal K$}}
\put(400,1866){\makebox(0,0)[r]{$L_\infty$}}
\end{picture}

\normalsize
\noindent
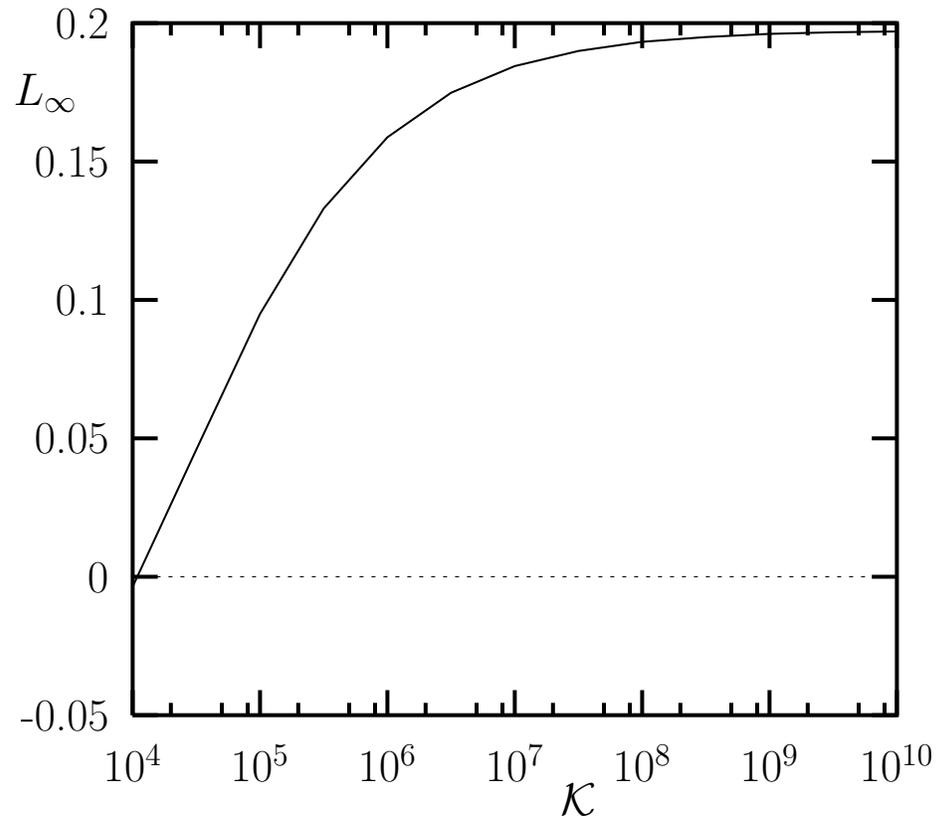
{\bf Figure 1} \\
Numerical solution of $L(\kappa \! \to \! \infty)$ as defined in
Equation (\ref{Lkappa}) using the full flow equations
(\ref{flow}). The potential strength is $U_0 = 0.1$; for large values
of ${\cal K} \! \ge \! 10^8$ the value $2 \pi^2 U_0^2$ (\ref{Linfty})
is attained.

\end{document}